\documentclass{llncs}

\usepackage{amssymb,amsmath,amstext,url,graphicx}

\begin{document}
\frontmatter          
\pagestyle{headings}  

\title{
	Estimating the dimensionality of neural responses with fMRI Repetition Suppression
}

\author{
	Mattia Rigotti\inst{1} \and 
	Stefano Fusi\inst{2}
}

\institute{
	IBM T.J.\ Watson Research Center, Yorktown Heights, NY 10598, USA
	\and Center for Theoretical Neuroscience, Department of Neuroscience, Columbia University Medical Center, New York, NY 10032, USA
}

\maketitle

\begin{abstract}
We propose a novel method that exploits fMRI Repetition Suppression (RS-fMRI) to measure the dimensionality of the set response vectors, i.e.\ the dimension of the space of linear combinations of neural population activity patterns in response to specific task conditions. RS-fMRI measures the overlap between response vectors even in brain areas displaying no discernible average differential BOLD signal. We show how this property can be used to estimate the neural response dimensionality in areas lacking macroscopic spatial patterning. The importance of dimensionality derives from how it relates to a neural circuit's functionality. As we show, the dimensionality of the response vectors is predicted to be high in areas involved in multi-stream integration, while it is low in areas where inputs from independent sources do not interact or merely overlap linearly. Our method can be used to identify and functionally characterize cortical circuits that integrate multiple independent information pathways.
\end{abstract}

\def\beq{\begin{eqnarray}}
\def\eeq{\end{eqnarray}}

\renewcommand{\vec}[1]{\mathbf{\ensuremath{#1}}}
\newcommand{\gvec}[1]{\ensuremath{\boldsymbol{#1}}} 

\def\wb{\vec{w}}
\def\wt{\vec{\widetilde{w}}}
\def\hb{\vec{h}}
\def\wbt{\vec{\widetilde{w}}}
\def\hbt{\vec{\widetilde{h}}}
\def\wbh{\vec{\widehat{w}}}
\def\hbh{\vec{\widehat{h}}}
\def\kt{\widetilde{\kappa}}
\def\kh{\hat{\kappa}}

\def\LMD{\Sigma}

\section{Introduction}

The pioneering work by Hubel and Wiesel \cite{Hubel1959} on the receptive field properties of neurons in early visual cortex has helped shape the remarkably influential notion that sensory information in the brain is processed in hierarchically organized computational modules. Feedforward stages where features are distinctly extracted and analyzed feed into alternating higher order convergent stages where such features are combined to form the increasingly complex receptive fields of more specialized cells \cite{Barlow1972}.

The idea to use multiple hierarchical elaboration stages of increasing levels of abstraction has been embraced by the deep learning community with spectacular results in processing tasks of images, audio and video \cite{LeCun2015}. However, the general question as to how higher order features should be derived from the convergence of more basic ones is still unresolved, and has long been at the center of a multidisciplinary interest that spans computational and systems neuroscience, computer vision, and machine learning in general. One central idea from this interdisciplinary cross-pollination is that higher order features have to strike a balance between \emph{dimensionality expansion} and \emph{dimensionality reduction}. The convergence of lower order features within higher order modules has to be non-linear, so as to guarantee dimensionality expansion and therefore high-margin separation \cite{Barak2011,Barak2013}. On the other hand, non-linear dimensionality expansion is usually alternated with a dimensionality reduction step, such as max-pooling, which allows for better generalization by enforcing invariance to input changes that shouldn't affect the final processing output \cite{DiCarlo2007,Anselmi2013,Wang2014}.
From a kernel machine perspective, good resulting high order features should be rich enough to serve as a ``basis'' to express the final output, while at the same time allowing for a parsimonious expression in terms of a reduced number of components \cite{Montavon2011}. This complements the emerging view in the deep learning community that in the higher layers of deep neural networks trained to perform classification tasks, the important semantic information is contained in the space of linear combinations of responses, rather than in the activity of individual units \cite{Szegedy2013}.

Going beyond sensory processing tasks, recent neuroscience studies have confirmed the importance of the specific structure of the linear space generated by the neural responses also during the execution of cognitive operations. In particular, the dimensionality of the response vectors in convergent frontal cortical areas has been shown to predict the behavior of rodents \cite{Balaguer-Ballester2011} and monkeys \cite{Rigotti2013} during working memory tasks, and to collapse during error trials.

In this contribution we propose a way to quantify the dimensionality of the neural response patterns using repetition suppression fMRI (RS-fMRI). RS-fMRI or fMRI-adaptation \cite{Grill-Spector2006,Grill-Spector2001} refers to a progressive reduction of BOLD response following repeated stimulus presentations. The subsequent presentation of differing stimuli results in a BOLD signal suppression that can be used to estimate the extent to which their representations overlap at the neural level. We propose to use this approach to determine how dissimilar responses are within a single voxel. This approach enables us to measure the relative distances between neural responses and produce an estimate of the dimensionality within a voxel. 

As we will show in the Results section, this method can be used to investigate the role of specific brain regions in dimensionality expansion or reduction,  potentially contributing to the understanding of the computational and algorithmic basis of higher order cognitive functions. As an application, we will illustrate the result of simulations indicating that our method can be used to identify and functionally characterize areas that integrate multiple information pathways.

\section{Methods}

\subsection{Model of repetition suppression in fMRI}

This section illustrates in detail the assumptions of our RS-fMRI model, and demonstrates how this technique can be leveraged to estimate the dimensionality of the response vectors within voxels.

\vspace{-0.5cm}
\subsubsection{Additive model of fMRI BOLD signal from neural activity.}

We denote with $\mathbf{f}(e)$ the response vector elicited in response to a task-relevant event $e$ within a brain region or voxel of interest of $N$ neurons. Accordingly, the firing rate activity of neuron $i$ during event $e$ is indicated by $f_i(e)$. We are interested in the dimensionality of the space spanned by the vectors  $\mathbf{f}(e_k)$, $k=1,\ldots,n$. This quantity equals the rank of the matrix $F = \bigl(\mathbf{f}(e_1), \quad\mathbf{f}(e_2), ~\ldots~\mathbf{f}(e_n)\bigr)$ obtained by concatenating all response vectors.

We model the fMRI signal in the region of interest as an aggregate function of the response function of all $N$ neurons. In particular, we assume that the BOLD signal evoked by event $e$ follows the simple form: 
\begin{equation}
B_e  = \beta \sum_i^N f_i(e),
\label{eq:BOLD}
\end{equation} 
where $\beta$ is a positive constant relating neural activation to BOLD signal. Notice that $B_e$ is a function of the population response vector evoked by event $e$, but is a scalar. Previous studies coupling electrode recordings and intrinsic optical imaging of BOLD signals demonstrate the plausibility of assuming such a linear relation between BOLD and spiking activity \cite{Cardoso2012}. In all simulations we add a Gaussian random variable to $B_e$ to simulate fMRI noise \cite{Kruger2001}.

\vspace{-0.5cm}
\subsubsection{Firing rate adaptation model of RS-fMRI.}

We mechanistically model repetition suppression as being due to firing rate adaptation or ``neuronal fatigue'' \cite{Grill-Spector2006,DeBaene2010}. We set out to model the neural response elicited by the presentation of sequences of two events $e_1$ and $e_2$ in a fixed temporal order. In particular, we model the fact that the response vector at the presentation of the second event will be modified, due to the previous activation at the first event. Our model assumes that the firing rate response of neuron $i$ at the first event $e_1$ is given by
$f_i(e_1)=\phi(I_i(e_1))$,
where $I_i(e_1)$ indicates the total synaptic current to neuron $i$ and $\phi(\cdot)$ indicates the neuronal current-to-firing rate transfer function.

We will use the model of firing rate adaptation developed by \cite{LaCamera2004}, and assume that the firing response activates an hyperpolarizing current proportional to the response itself. More specifically, at the presentation of a second event $e_2$, following a fixed delay after event $e_1$, we model the response of neuron $i$ as
\begin{equation}
    f_i(e_1e_2)=\phi\left(I_i(e_2)-\alpha_t f_i(e_1)\right),
    \label{eq:f_i(e_1e_2)}
\end{equation}
where $I_i(e_2)$ is the synaptic drive elicited by event $e_2$ onto neuron $i$, $\alpha_t$ is a positive parameter that quantifies the adaptation strength (the subscript $t$ denotes the dependence of the adaptation effect to the time interval between $e_1$ and $e_2$), and $f_i(e_1)$ is the unadapted response to event $e_1$.

\vspace{-0.5cm}
\subsubsection{RS-fMRI and neural response vectors dimensionality.}

As it is often done, we assume that the neurons have a transfer function that goes asymptotically to zero for decreasing inputs and tends asymptotically to a linear function for growing inputs \cite{Abbott2005}. In-between, we assume that the transfer function can be modeled as an exponential: $\phi(I)=\exp(I/\sigma+b)$, with some constant $\sigma$ and a bias term $b$. In this regime, equation \eqref{eq:f_i(e_1e_2)} assumes the multiplicative form:
\begin{equation*}
	f_i(e_1e_2)=\phi\left(I_i(e_2)-\alpha_t f_i(e_1)\right)=\rho\left(f_i(e_1)\right)f_i(e_2),
\end{equation*}
with $\rho(x) = \exp(-\alpha_t x/\sigma)$, resulting in a fMRI-RS signal described by:
\begin{equation}
	B_{e_1 e_2} = \beta \sum_i \rho(f_i(e_1))f_i(e_2).
	\label{eq:BOLD_RS}
\end{equation}
Notice that the factors $\rho(f_i(e_1))$ are implicitly also a function of the inter-event interval between $e_1$ and $e_2$ through the dependence on $\alpha_t$. In the limit where the two events are very distant in time, their value should in fact tend to $1$, and $B_{e_1 e_2}$ tend to $B_{e_2}$. In that limit, we can expand $\rho(\cdot)$ to first order as $\rho(f) \approx 1-\alpha_t f/\sigma$. We now define the Repetitions Suppression (RS) matrix with components $R_{kl}$:
\begin{align}
	R_{kl}&=B_{e_l}-B_{e_k e_l}
	\label{eq:def_R_k_l}
\end{align}
and notice that plugging the approximation $\rho(f) = 1-\alpha_t f/\sigma$ into eq.\ \eqref{eq:BOLD_RS} together with eq.\ \eqref{eq:BOLD} allows us to derive
\begin{equation*}
	R_{kl}=B_{e_l}-B_{e_k e_l} \propto \sum_i^N f_i(e_k)f_i(e_l)=(F^\top F)_{kl},
\end{equation*}
where $F$ is the previously defined matrix of response vectors. Now, because in general $rank(F) = rank(F^\top F)$, we get that the rank of the response vectors $F$ is the same as the rank of the RS-matrix $R_{kl}$.

Going through the same calculation in the case of a linear transfer function, it is easy to see that the constant and the asymptotically linear regime of the transfer function does not contribute to the rank. So, all contributions are given by the non-linear exponential part. Notice also that a threshold linear function (the transfer function we will use in the simulation) can be approximated by a piece-wise exponential and linear function.

As a final note it is important to mention that the estimate of the rank of a matrix is made notoriously difficult by random noise fluctuations. These random variations bias the estimate of the principal components, increasing the number of non-zero components that contribute to the rank (see e.g.\ \cite{Sengupta1999}).

We determine the noise floor similarly to \cite{Machens2010} by estimating a noise covariance matrix from single-trial data that we use to build a distribution of ``noise eigenvalues''. We then set the noise threshold to $95\%$ of the maximal noise eigenvalue. In all simulations we set a noise level that corresponds to a SNR of 100, a value that matches measures of physiological noise in gray matter \cite{Kruger2001}.

\section{Results}

\subsection{RS-matrix estimate of dimensionality of response vectors}

\begin{figure}
\centering
\includegraphics[width=0.9\linewidth]{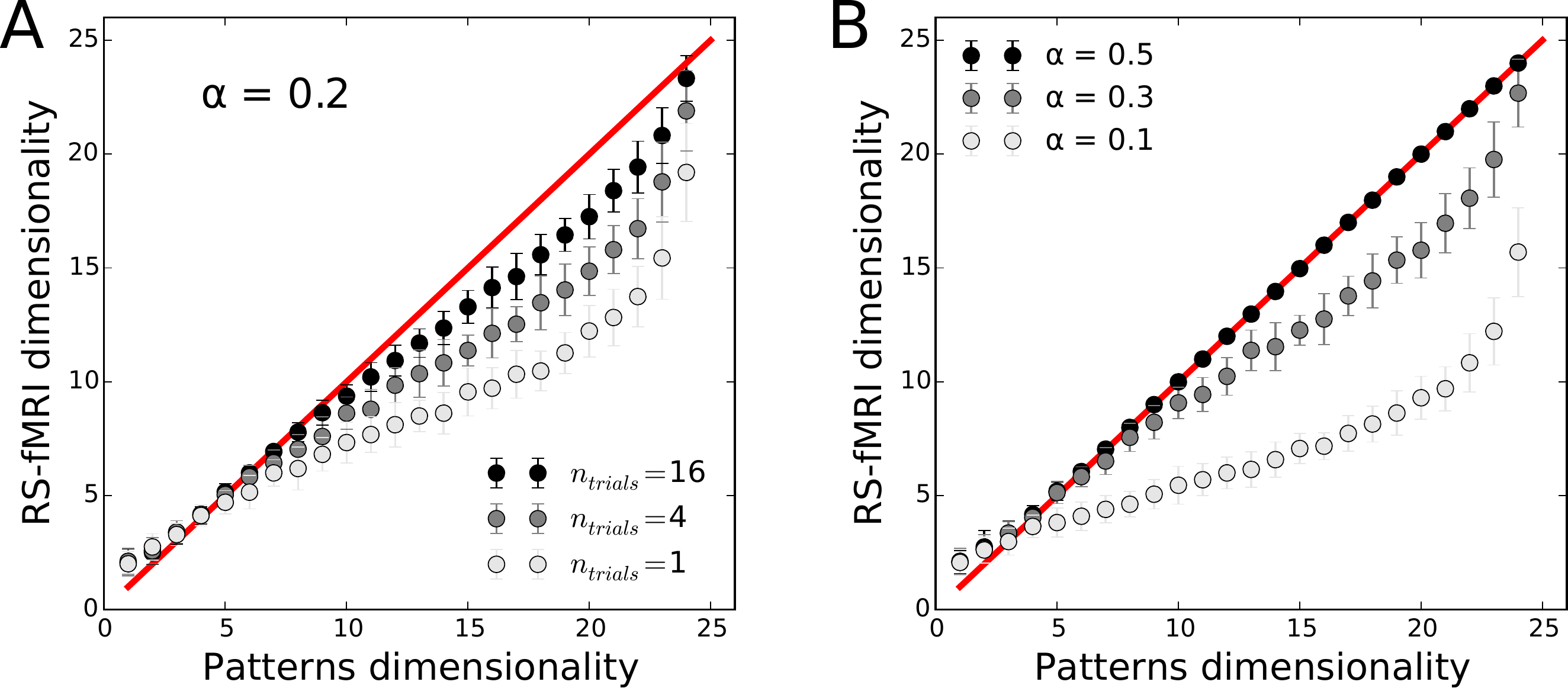}
\caption{Estimate obtained from the RS-matrix accurately quantifies the dimensionality of the neural response vectors. \textbf{A} Dimensionality estimated from the RS-matrix is highly correlated with the actual patterns dimensionality. Moreover, the correlation grows as more and more trials are used to average out noise in the RS-fMRI estimate. The plot shows the RS-fMRI estimated dimensionality as a function of actual dimensionality with no trial-to-trial averaging (white circles), average over 4 trials (gray circles) and average over 16 trials (black circles) for a fixed number of $p=25$ patterns with rank growing between 1 and 25. The circles indicate the mean over 100 simulations, while error bars denote the corresponding standard deviations. \textbf{B} As in \textbf{A}, but with no trial-to-trial average (only 1 trial) and simulating different values of the adaptation factor $\alpha$.}
\label{fig:fig_correlation}
\end{figure}

The following simulations validate our method by showing that our model of BOLD response eq.\ \eqref{eq:BOLD} with adapting neurons eq.\ \eqref{eq:f_i(e_1e_2)} allows us to reliably estimate the dimensionality of the response vectors from the RS-matrix $R$ in eq.\ \eqref{eq:def_R_k_l}. We generate sets of $p$ $N$-dimensional patterns with varying rank by multiplying an $N\times d$-dimensional random Gaussian matrix (centered at zero and with unitary variance) with a  $d\times p$ random binary matrix. These vectors correspond to the synaptic currents $I_i(e)$ generating the firing rates $f_i(e)=\phi(I_i(e))$. As a transfer function $\phi(\cdot)$ we use a threshold-linear function $\phi(I)=\max(I,0)$.  We then use eq.\ \eqref{eq:f_i(e_1e_2)} with eq.\ \eqref{eq:def_R_k_l} and \eqref{eq:BOLD} to generate the RS-matrix $R$. For numerical stability reasons, we then symmetrize this matrix by averaging it with its own transpose. We then estimate the dimensionality by comparing the eigenvalues of this matrix to the noise threshold as explained in the Methods section and compare it with the dimensionality obtained from the matrix $F_{ik}=f_i(e_k)$. Figure \ref{fig:fig_correlation} shows the result of such a simulation, demonstrating that the estimated dimensionality accurately correlates with the actual response vectors dimensionality.

\subsection{Functional dissection of information processing pathways}

\begin{figure}
\centering
\includegraphics[width=0.9\linewidth]{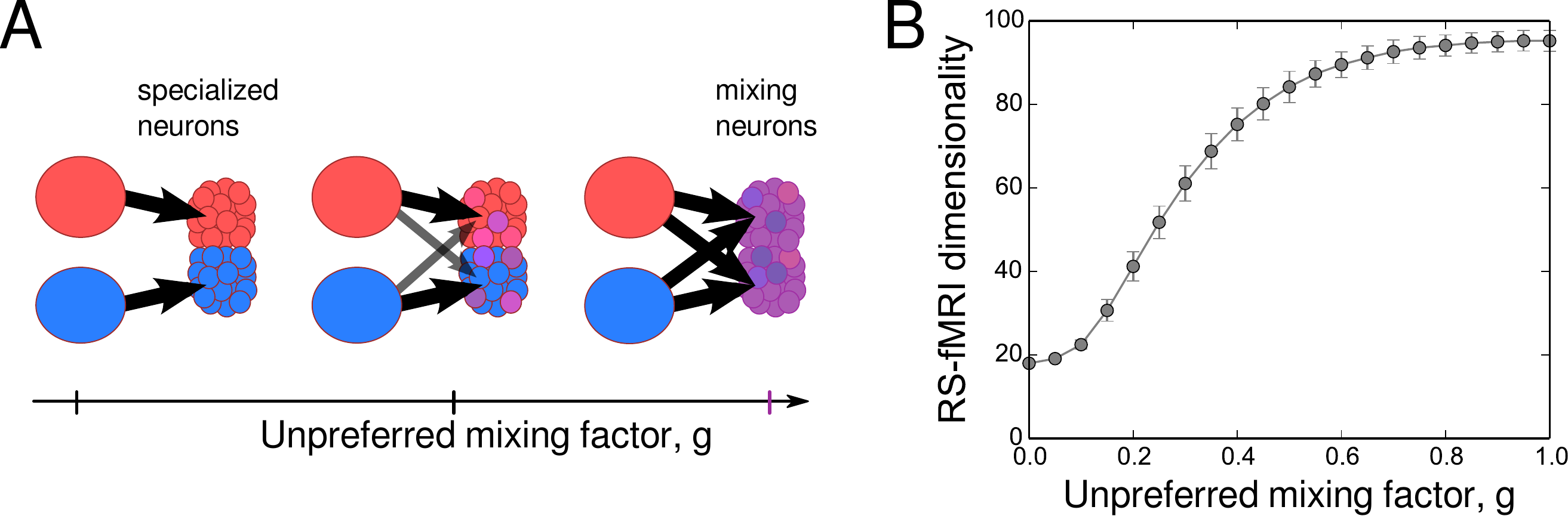}
\caption{Functional dissection of convergent pathways with RS-fMRI. \textbf{A} Scheme of the simulated circuit. The input areas project to a convergence area. The way the inputs are mixed in parametrized by $g$ (between 0 and 1). For $g=0$ every neuron in the convergence area receives inputs only from its preferred input area. As $g$ increases, the input from the unpreferred area increases until $g=1$ where the neurons mix on average equal contributions from both input areas. \textbf{B} Measuring the dimensionality of the response vectors in the convergence area as in Fig.\ \ref{fig:fig_correlation} shows, as expected (see \cite{Barak2013}), that the dimensionality grows as the mixing of the two inputs increases.}
\label{fig:dissection}
\end{figure}

We now illustrate how our technique could be used for the functional dissection of neural information processing in the brain. We imagine a neural circuit that receives synaptic projections from two separate independent input areas $A$ and $B$, each of which can be in $10$ states indicised by $k_A$ and $k_B$, respectively. In our model every neuron has a preferred input (either $A$ or $B$) from which it receives most of its inputs. So, a neuron $i$ with a preference for $A$ will receive a current $I_i(k_A,k_B)=\sum_j w^A_{ij}~f_j^{A}(k_A) + g \sum_j w^B_{ij}~f_j^{B}(k_B)$, where the ``unpreferred mixing factor'' $g$ is a number from $0$ and $1$ that quantifies the contribution of the unpreferred input, $w^A$ and $w^B$ are random centered Gaussian matrices with unitary variance, while $f_j^{A}(k_A)$ and $f_j^{B}(k_B)$ are the activities of the neurons in the two input areas (also taken as centered random Gaussian variables with unitary variance in our simulations). Everything works analogously for neurons with a $B$ preference. With $g=0$ the neurons in the convergence area will only receive projections from their preferred input, while for $g=1$ the neurons will equally mix contribution from both inputs (see Fig.\ \ref{fig:dissection}A). Panel Fig.\ \ref{fig:dissection}B shows the results of the simulation carried out similarly to the one in the previous section to simulate the process of estimating the dimensionality of the responses in the convergence areas. If the neurons in the convergence area only receive input from their preferred area the dimensionality is estimated to be $19=20-1$, the result of separately counting the contribution of the two areas (see \cite{Barak2013}). As the mixing factor increases and neurons mix more and more the contributions from the two inputs, the measured dimensionality increases until it saturates around $99=100-1=10\times 10-1$ when the two inputs are equally mixed (see \cite{Barak2013}). This could be used to identify areas that receive contributions from multiple independent inputs, and to characterize the type of functional convergence.

\section{Discussion}

We proposed a novel technique for designing experiments and analyzing data that will allow us to estimate dimensionality in single voxels using fMRI. If our mechanistic model is accurate, the dimensionality estimate given by our method should match the result from single unit recordings, in the sense that having access to the individual activity of all neurons in the voxel wouldn't give any advantage. This is because of the mathematical identity relating the rank of the response vectors and that of their overlaps captured by the RS-matrix. Here we put forth the main idea with analytical calculations on realistic mechanistic models of repetition suppression and we showed two simulations illustrating possible applications. Further simulations (omitted because of space constraints) show that our technique is robust to variations of the RS model (e.g.\ the inclusion of synaptic effects such as ``input fatigue''), to different choices of non-linear transfer functions, forms of noise, and heterogeneity of RS parameters. 

Further ongoing studies will compare the efficacy of this new technique to alternative methods for estimating dimensionality based on Multi-Voxel Pattern Analysis (MVPA) \cite{Norman2006}. Crucially, MVPA relies on a differential signal within individual voxels, while voxels whose BOLD activity is not modulated across task conditions, won't contribute to decoding. It is important to note that this situation where a voxel does not contribute to the differential BOLD signal can arise in two distinct mechanistic cases: 1) the trivial scenario where the activity of individual neurons within the voxel don't display any selectivity to task conditions, and 2) the case where neurons are indeed selective to specific conditions, but the preferences across neurons within the voxel are balanced in such a way that the signal is averaged out when it is aggregated in the total BOLD activity. In the second case, RS-fMRI can still distinguish between conditions, since the activity patterns are effectively distinct.

We therefore expect MVPA to be maximally effective when voxels are organized in such a way as to display a net preference, due to an underlying preference similarity (which can be quantified as signal correlation \cite{Averbeck2006}) across its constituting neurons, while we expect it to be progressively less effective as the preference of neurons in the voxels is less and less correlated, until the extreme case where it completely fails due to preferences being perfectly balanced. Notice that the situation in which we are interested is one where no particular anatomical organization is present, such as it is observed in higher order cortices, and therefore where we expect RS-fMRI to have a competitive advantage over MVPA.

These intuitive considerations are borne out quantitatively by a model of anatomical organization that shows that in the limit of disappearing signal correlation between neurons in a voxel, the MVPA signal tends to a small term that scales as $1/\sqrt{N}$, where $N$ is the number of neurons in the voxel (a finite-size effect signal), while the RS-fMRI signal remains of the same order as with finite signal correlation \cite{Rigotti_inprep}.

Clearly, our method remains to be demonstrated empirically. In this respect we're encouraged by previous studies showing that RS-fMRI can be utilized to distinguish between cortical areas tuned to either independent or conjoint visual features \cite{Drucker2009}. A conclusive empirical verification of our technique for the measure of response dimensionality is however left to future experiments.

\section{Acknowledgements}
SF was supported by the Gatsby Charitable Foundation, the Simons Foundation, the Swartz Foundation, the Kavli Foundation and the Grossman Foundation.

\bibliographystyle{splncs03}
\bibliography{RS_nips}

\end{document}